\documentclass[a4paper,11pt]{article}
\pdfoutput=1 

\usepackage{jcappub} 

\usepackage[T1]{fontenc} 

\usepackage{aas_macros}

\usepackage{lineno}
\usepackage{bm}
\usepackage{tabularx}

\title{Search for heavy dark matter from dwarf spheroidal galaxies: leveraging cascades and subhalo models}

\author[a]{Deheng Song}
\author[b,c,d]{Nagisa Hiroshima}
\author[e,f,g,a]{Kohta Murase}

\affiliation[a]{Center for Gravitational Physics and Quantum Information,
Yukawa Institute for Theoretical Physics, Kyoto University, Kyoto 606-8502, Japan}
\affiliation[b]{Department of Physics, Faculty of Engineering Science, Yokohama National University, Yokohama 240–8501, Japan}
\affiliation[c]{Department of Physics, University of Toyama, 3190 Gofuku, Toyama 930-8555, Japan}
\affiliation[d]{RIKEN Interdisciplinary Theoretical and Mathematical Sciences (iTHEMS),
Wako, Saitama 351-0198, Japan}
\affiliation[e]{Department of Physics, The Pennsylvania State University, University Park, Pennsylvania 16802, USA}
\affiliation[f]{Department of Astronomy and Astrophysics, The Pennsylvania State University, University Park, Pennsylvania 16802, USA}
\affiliation[g]{Center for Multimessenger Astrophysics, The Pennsylvania State University, University Park, Pennsylvania 16802, USA}

\emailAdd{songdeheng@yukawa.kyoto-u.ac.jp}
\emailAdd{hiroshima-nagisa-hd@ynu.ac.jp}
\emailAdd{murase@psu.edu}

\abstract{
The {\it Fermi} Large Area Telescope ({\it Fermi}-LAT) has been widely used to search for Weakly Interacting Massive Particle (WIMP) dark matter signals due to its unparalleled sensitivity in the GeV energy band. The leading constraints for WIMP by {\it Fermi}-LAT are obtained from the analyses of dwarf spheroidal galaxies within the Local Group, which are compelling targets for dark matter searches due to their relatively low astrophysical backgrounds and high dark matter content. In the meantime, the search for heavy dark matter with masses above TeV remains a compelling and relatively unexplored frontier. In this study, we utilize 14-year {\it Fermi}-LAT data to search for dark matter annihilation and decay signals in 8 classical dwarf spheroidal galaxies within the Local Group. We consider secondary emission caused by electromagnetic cascades of prompt gamma rays and electrons/positrons from dark matter, which enables us to extend the search with {\it Fermi}-LAT to heavier dark matter cases. We also update the dark matter subhalo model with informative priors respecting the fact that they reside in subhalos of our Milky Way halo aiming to enhance the robustness of our results. We place constraints on dark matter annihilation cross section and decay lifetime for dark matter masses ranging from $10^3$ GeV to $10^{11}$ GeV, where our limits are more stringent than those obtained by many other high-energy gamma-ray instruments.}

\begin{document}
\setpagewiselinenumbers
\maketitle
\flushbottom

\section{Introduction}
\label{s:intro}
The nature of dark matter (DM) remains a mystery. Cosmological observations have shaped our understanding of the Universe, indicating that non-baryonic matter makes up approximately a quarter of the total energy density of the Universe~\cite{Aghanim:2018eyx,Akrami:2018odb}, thus giving rise to the DM problem. In the realm of particle physics, various candidates have been proposed to extend the Standard Model, and these candidates are currently under investigation through a combination of collider experiments, direct detection experiments, and astrophysical or cosmological observations (see Refs.~\cite{Bertone:2016nfn,Lin:2019uvt,Safdi:2022xkm} for recent reviews). For example, the Weakly Interacting Massive Particle (WIMP), which has long been a leading candidate, is already tightly constrained from multiple aspects for masses below approximately $m\lesssim{\cal O}(100)$~GeV ~\cite{Hoof:2018hyn,GAMBIT:2021rlp}. 

In this article, we focus on DM heavier than WIMPs, with masses above approximately 1 TeV. For such heavy DM, its relic abundance is not necessarily determined by thermal freeze-out~\cite{Srednicki:1988ce, Dubrovich:2003jg, Hisano:2006nn, Steigman:2012nb, vonHarling:2014kha, Bramante:2017obj, Baldes:2017gzw, Cirelli:2018iax, Smirnov:2019ngs, Bhatia:2020itt,  Baldes:2021aph}. In this case, collider experiments face kinematic limitations in creation processes, and the scattering rate with underground detectors decreases as the number density of incoming DM particles decreases. 
For this reason, high-energy astrophysical observation plays a crucial role {because it provides a unique window for probing heavy DM.}
Many of {current} constraints on the annihilation cross sections and lifetime of heavy DM are based on observations of electromagnetic emission, which use various targets with distinct advantages to probe DM include the diffuse gamma-ray background~\cite{Abazajian:2010zb, Murase:2012xs, Bringmann:2013ruh,Ajello:2015mfa, Fermi-LAT:2015qzw, DiMauro:2015ika, DiMauro:2015tfa, Liu:2016ngs, Blanco:2018esa}, galaxy clusters~\cite{2010JCAP...05..025A, Huang:2011xr, 2011PhRvD..84l3509P,2012ApJ...750..123A,2012JCAP...07..017A, Murase:2012rd, Urban:2014yda,Fermi-LAT:2015xij, Tan:2019gmb, Thorpe-Morgan:2020czg, DiMauro:2023qat, Song:2023xdk}, the Milky Way halo~\cite{Bergstrom:1997fj, Ascasibar:2005rw, 2010PhRvD..81d3532K, Murase:2015gea,Cohen:2016uyg,Chang:2018bpt,Maity:2021umk, Tak:2022vkb}, and dwarf spheroidal galaxies (dSphs)~\cite{Stoehr:2003hf, Diemand:2006ik, Cholis:2012am, Carlson:2014nra, Lopez:2015uma, Fermi-LAT:2015att, Bonnivard:2015xpq,Baring:2015sza, MAGIC:2016xys, Calore:2018sdx, Ando:2019rvr, Hoof:2018hyn, Kar:2019cqo, Alvarez:2020cmw, Gammaldi:2021zdm, Ando:2021fhj}. Multi-messenger approaches are also powerful, and other constraints on heavy DM include those from charged cosmic rays~\cite{Yoshida:1998it, Blasi:2000ud, Marzola:2016hyt, Cuoco:2017iax, Supanitsky:2019ayx, Alcantara:2019sco, Ishiwata:2019aet, Motz:2023mez, Das:2023wtk},  neutrinos~\cite{Murase:2012rd,IceCube:2013bas,Murase:2015gea,Kar:2019cqo,Ishiwata:2019aet,Chianese:2021htv,Fiorillo:2023clw}, {and the cosmic microwave background (CMB)~\cite{Slatyer:2009yq, Kawasaki:2021etm}}. Among these various indirect searches for heavy DM, dSphs are promising targets. The kinematics of stars in dSphs, which are satellite galaxies in subhalos of our Galaxy, indicate that they hold a huge amount of DM. However, the detailed DM density profile of each target is still uncertain and this dominates the uncertainty of the current limits for DM derived with dSphs. The nature of dSphs in subhalos that they have experienced tidal disruption under the potential of the Milky Way makes it difficult to obtain precise estimates.

One specific nature of heavy DM of $m_\chi\gtrsim{\cal O}(1)$~TeV is that the products from DM annihilation/decay, i.e. such as prompt gamma rays and electrons/positrons, interact with background photon fields and magnetic fields before reaching Earth. Prompt emission, occurring on the scale of the DM mass, may cascade down to lower-energy levels through electromagnetic cascades (e.g., Refs.~\cite{Murase:2012xs,Murase:2012rd}). In other words, heavy DM with much greater masses can still leave its signature in the energy range of GeV or even lower. Therefore, heavy DM can be probed by instruments like the {\it Fermi} Large Area Telescope ({\it Fermi}-LAT).

In this work, we investigate annihilation and decay signatures of heavy DM in 8 classical dSphs using 14-year {\it Fermi}-LAT data. These dSphs are nearby and their properties are relatively well probed. The secondary gamma-ray emission resulting from electromagnetic cascades of prompt gamma rays and electrons/positrons originating from heavy DM with masses exceeding 1~TeV are calculated based on Ref.~\cite{Murase:2012rd} with the one-zone approximation. We incorporate the spatial extension of the target dSphs which retain the features of tidal interaction by constructing the emission template based on Ref.~\cite{Ando:2020yyk}. By performing the profile likelihood analysis, we constrain the annihilation cross section and decay lifetime of heavy DM. 

The structure of the paper is as follows. In section~\ref{s:shmodel}, we describe the subhalo model for dSphs. In section~\ref{s:dm_model}, we explain the expected signals from heavy DM, including electromagnetic cascades in dSphs. We detail our data analysis in section~\ref{s:analysis}. In section~\ref{s:results}, we present and discuss our results, and we conclude the paper in section~\ref{s:conclusion}.

\section{Subhalo model}
\label{s:shmodel}
The challenge of precisely modeling the DM density distribution in dSphs is well-recognized, and various models have been proposed in the literature  (e.g. see Refs.~\cite{Martinez:2013els,Geringer-Sameth:2014yza,Hayashi:2016kcy,Pace:2018tin}). In this work, we adopt the model proposed in Ref.~\cite{Ando:2020yyk} which applies informative prior respecting the fact that dSphs reside in subhalos of the Milky Way to improve the accuracy of parameter estimates for the DM density profile.

The DM density distribution is described using the Navarro-Frenk-White (NFW) profile~\cite{Navarro:1996gj} with truncation:
\begin{equation}
    \rho_\chi(r)=\left\{\begin{array}{ll}
         \rho_s\left(\frac{r}{r_s}\right)^{-1}\left(1+\frac{r}{r_s}\right)^{-2}, & r<r_t, \\
         0,& r \geq r_t.
    \end{array}\right.
\end{equation}
The profile is characterized by two parameters, $\rho_s$ and $r_s$, for the NFW model and the truncation radius $r_t$. Likelihood analysis of stellar kinematics data results in degenerate constraints on the $\rho_s$--$r_s$ plane. In this work, we consider the following 8 classical dSphs: Carina~\cite{Walker:2008fc}, Draco~\cite{Walker:2015,Walker:2009zp}, Fornax~\cite{Walker:2008fc}, Leo~I~\cite{Mateo:2007xh}, Leo~II~\cite{Spencer:2017}, Sculptor~\cite{Walker:2008fc}, Sextans~\cite{Walker:2008fc}, and Ursa Minor~\cite{Walker:2009zp}. We exclude Sagittarius from our analysis, as observations indicate that this dSph is currently undergoing disruption~\cite{Mateo:1996}, which could introduce significant uncertainty in the profile parameters. The distances and Galactic coordinates of these dSphs are listed in Table~\ref{tab:dsphs}.
\begin{table}[t!]
    \centering
    \begin{tabular}{cccc}
    \hline
Name & Distance & $l$   & $b$ \\
     & [kpc]    & [deg] & [deg]  \\\hline
Carina & 105.0 $\pm$ 6.0 & 260.11 & -22.22 \\
Draco & 76.0 $\pm$ 6.0 & 86.37 & 34.72 \\
Fornax & 147.0 $\pm$ 12.0 & 237.10 & -65.65 \\
Leo I & 254.0 $\pm$ 15.0 & 225.99 & 49.11 \\
Leo II & 233.0 $\pm$ 14.0 & 220.17 & 67.23 \\
Sculptor & 86.0 $\pm$ 6.0 & 287.54 & -83.16 \\
Sextans & 86.0 $\pm$ 4.0 & 243.50 & 42.27 \\
Ursa Minor & 76.0 $\pm$ 3.0 & 104.97 & 44.80 \\
\hline
    \end{tabular}
    \caption{Distances and Galactic coordinates of 8 classical dSphs.}
    \label{tab:dsphs}
\end{table}

Since dSphs reside in subhalos of our Galaxy, they should experience tidal stripping by the Milky Way, leading to the current diversity in their DM density profiles. We take the Bayesian approach of Ref.~\cite{Ando:2020yyk} to reduce the uncertainties in the $\rho_s$, $r_s$, and $r_t$.  The prior distribution suitable for each target dSph is generated using the model for subhalo evolutions~\cite{Hiroshima:2018kfv}. The mass and the redshift distribution of accreting subhalos to the Milky Way is evaluated using the Extended Press-Schechter model~\cite{Yang:2011rf}. The tidal mass-loss rate is evaluated at the pericenter of each accreted subhalo~\cite{Jiang:2014nsa}. The evolution of the DM density profile parameter is determined through the relationship derived in Ref.~\cite{Penarrubia:2010jk}{, which characterizes the evolution of the density profile parameters ($\rho_s, r_s$) as a function of the mass ratio before and after the tidal-mass loss. The parameter of the fitting function are calibrated against simulations.} For the subhalo-satellite connection, we adopt the model of Ref.~\cite{Graus:2019} with a threshold for the maximum circular velocity at accretion ($V_{\rm peak} > 25$ km/s) and a velocity dispersion of $\sigma = 2.5$ km/s as specified in eq.~2 of Ref.~\cite{Ando:2020yyk}. These criteria are particularly suitable for classical dSphs~\cite{Hargis:2014kaa}.

We model the angular extension of the target dSphs by assuming the median values of $\rho_s$, $r_s$, and $r_t$ obtained from the simulations. To further restrict the angular extension of the dSphs, we use the radii of the outermost stars~\cite{Geringer-Sameth:2014yza} in the target dSphs. Specifically, we calculate the angular extension {$\theta_\Delta$} of each dSph up to 
\begin{equation}
    \theta_\Delta = \sin^{-1}(r_\Delta/d_c),
\end{equation}
where $d_c$ is the comoving distance of the dSph and
\begin{equation}
    r_\Delta = \min(r_t, r_\mathrm{max}).
\end{equation}
{The hierarchy between the estimated $r_t$ and the observed radius of the outermost member star depends on the target dSph. So we introduce the quantity $r_\Delta$ for obtaining conservative J-factors.} We calculate the J-factors of the dSphs for annihilation/decay ($J^\mathrm{ann}$/$J^\mathrm{dec}$) up to $\theta_\Delta$ as follows,
\begin{equation}
    J^\mathrm{ann}(\theta_\Delta) = 2\pi \int_0^{\theta_\Delta} \sin{\theta}d\theta \int_\mathrm{l.o.s} dl \rho_\chi^2(r; \rho_s, r_s),
\end{equation}
and
\begin{equation}
    J^\mathrm{dec}(\theta_\Delta) = 2\pi \int_0^{\theta_\Delta} \sin{\theta}d\theta \int_\mathrm{l.o.s} dl \rho_\chi(r; \rho_s, r_s).
\end{equation}

\begin{figure}[ht!]
    \centering
    \includegraphics[width=0.8\columnwidth]{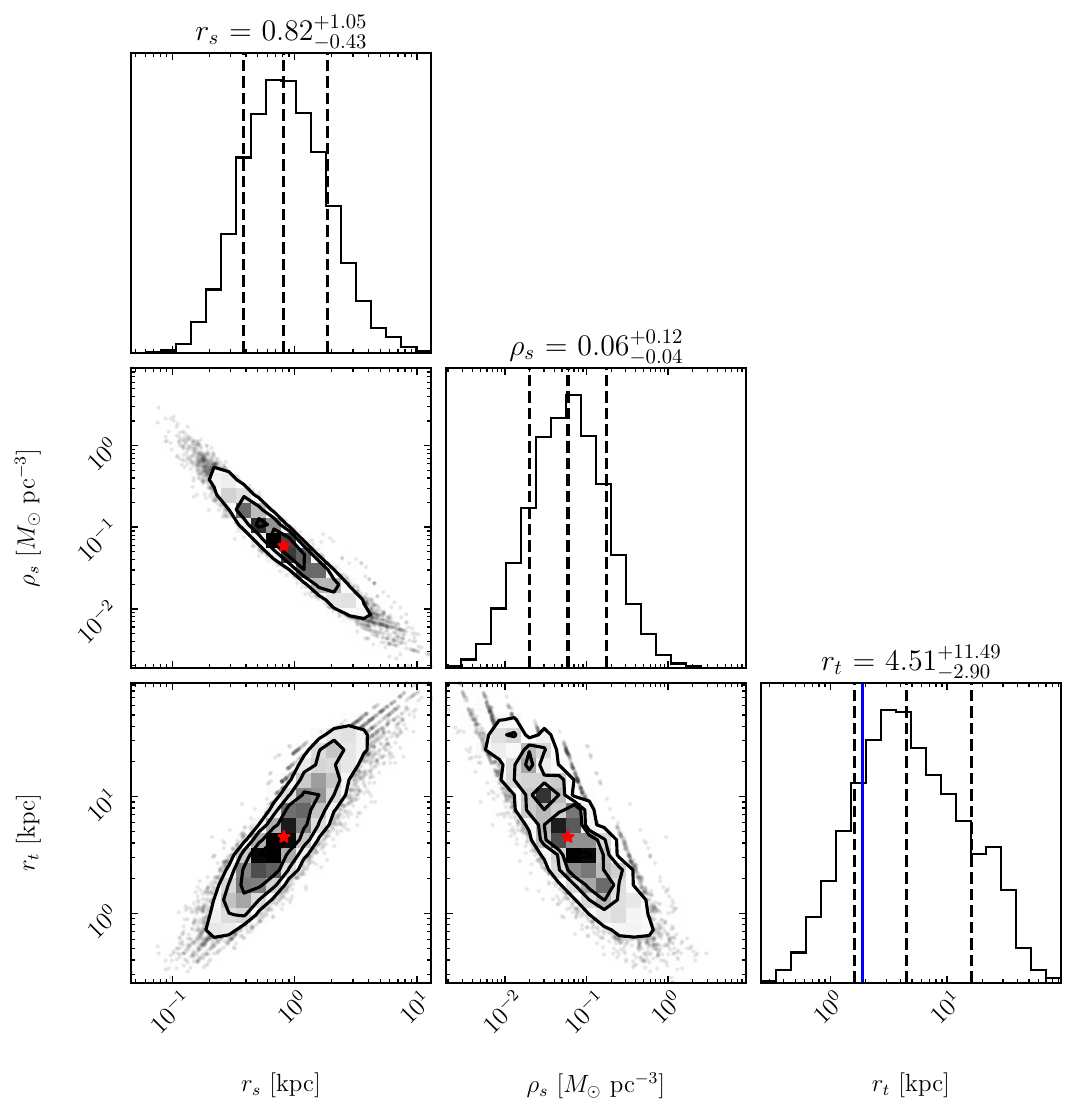}
    \caption{{Corner plot~\cite{corner} representing the distribution of $\rho_s$, $r_s$, and $r_t$ for Draco, using the model proposed in Ref.~\cite{Ando:2020yyk}. In the histograms, the median value and the 1-sigma percentile of the parameters are indicated by dashed vertical lines.} {The red stars represent the median values of the parameters. Here $r_\mathrm{max}$ of Draco is indicated by the blue solid line on the histogram of $r_t$.}}
    \label{fig:corner_Draco}
\end{figure}

\begin{figure}[ht!]
    \centering
    \includegraphics[width=0.48\columnwidth]{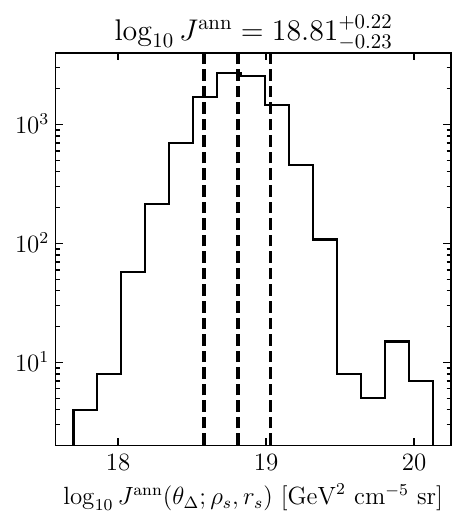}
    \includegraphics[width=0.48\columnwidth]{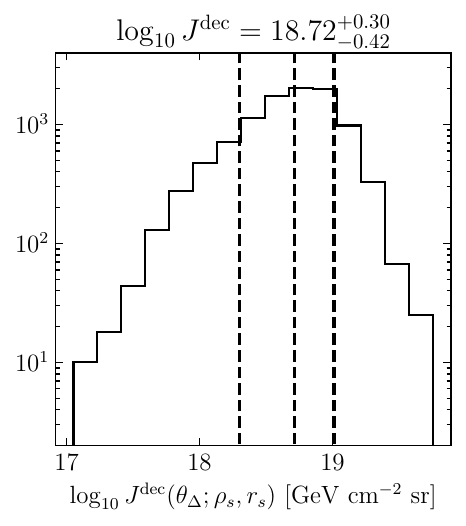}
    \caption{Histograms for $J^\mathrm{ann}(\theta_\Delta)$ and $J^\mathrm{dec}(\theta_\Delta)$ for Draco. {In the histograms, the median value and the 1-sigma percentile of the parameters are indicated by dashed vertical lines.}}
    \label{fig:hist_Draco}
\end{figure}

{In figure~\ref{fig:corner_Draco},  the corner plot of $\rho_s$, $r_s$, and $r_t$ parameters for Draco, obtained from the calculation based on the model proposed in Ref.~\cite{Ando:2020yyk}, is shown as example. The figure presents the 2-dimensional density plots between every pair of parameters, accompanied by 1-dimensional histograms for each parameter.} In the histograms, the median value and the 1-sigma percentile of the parameters are indicated by dashed vertical lines. Figure~\ref{fig:hist_Draco} shows the histograms for $J^\mathrm{ann}(\theta_\Delta)$ (left panel) and $J^\mathrm{dec}(\theta_\Delta)$ (right panel) of Draco, which are calculated based on the values of $\rho_s$, $r_s$, and $\theta_\Delta$ obtained from the simulations. The median values of $\rho_s$, $r_s$, and $r_t$ for the 8 target dSphs are listed in Table~\ref{tab:subhalos}. Additionally, Table~\ref{tab:subhalos} provides $r_\mathrm{max}$, $\theta_\Delta$, $J^\mathrm{ann}(\theta_\Delta)$, and $J^\mathrm{dec}(\theta_\Delta)$ for the target dSphs.
\begin{table}[ht!]
    \centering
    \bgroup
    \def\arraystretch{1.4}
    \addtolength{\tabcolsep}{-0.3em}
    \begin{tabularx}{\columnwidth}{cccccccc}\hline
    Name & ${\rho}_s$ & ${r}_s$ & ${r}_t$ & $r_\mathrm{max}$ & $\theta_\Delta$ & $\log_{10}J^\mathrm{ann}(\theta_\Delta)$ & $\log_{10}J^\mathrm{dec}(\theta_\Delta)$ \\
    & [$M_\odot$ pc$^{-3}$] & [kpc] & [kpc] & [kpc] & [deg] & [$\mathrm{GeV}^2\ \mathrm{cm}^{-5}$ sr] & [$\mathrm{GeV}\ \mathrm{cm}^{-2}$ sr] \\\hline
Carina & $0.12_{-0.08}^{+0.20}$ & $0.35_{-0.17}^{+0.45}$ & $1.29_{-0.68}^{+1.97}$ & $2.22_{-0.44}^{+0.89}$ & 0.70 & $18.02_{-0.26}^{+0.24}$ & $17.69_{-0.48}^{+0.56}$ \\
Draco & $0.06_{-0.04}^{+0.12}$ & $0.82_{-0.43}^{+1.05}$ & $4.51_{-2.90}^{+11.49}$ & $1.87_{-0.32}^{+0.72}$ & 1.41 & $18.81_{-0.23}^{+0.22}$ & $18.72_{-0.42}^{+0.30}$ \\
Fornax & $0.05_{-0.03}^{+0.08}$ & $0.96_{-0.42}^{+0.85}$ & $5.82_{-3.36}^{+9.57}$ & $6.27_{-1.37}^{+2.62}$ & 2.27 & $18.23_{-0.15}^{+0.17}$ & $18.46_{-0.38}^{+0.31}$ \\
Leo I & $0.05_{-0.03}^{+0.10}$ & $0.86_{-0.44}^{+0.95}$ & $4.97_{-3.06}^{+10.50}$ & $1.95_{-0.41}^{+0.79}$ & 0.44 & $17.73_{-0.08}^{+0.09}$ & $17.69_{-0.31}^{+0.23}$ \\
Leo II & $0.06_{-0.05}^{+0.15}$ & $0.62_{-0.33}^{+0.98}$ & $2.86_{-1.67}^{+7.03}$ & $0.82_{-0.18}^{+0.35}$ & 0.20 & $17.56_{-0.10}^{+0.10}$ & $17.22_{-0.24}^{+0.22}$ \\
Sculptor & $0.06_{-0.04}^{+0.11}$ & $0.80_{-0.42}^{+1.00}$ & $4.32_{-2.70}^{+10.75}$ & $2.67_{-0.57}^{+1.10}$ & 1.78 & $18.65_{-0.20}^{+0.19}$ & $18.67_{-0.49}^{+0.33}$ \\
Sextans & $0.09_{-0.06}^{+0.16}$ & $0.47_{-0.22}^{+0.56}$ & $1.93_{-1.05}^{+3.03}$ & $2.54_{-0.59}^{+1.11}$ & 1.28 & $18.35_{-0.25}^{+0.22}$ & $18.17_{-0.47}^{+0.46}$ \\
Ursa Minor & $0.06_{-0.04}^{+0.11}$ & $0.81_{-0.42}^{+0.97}$ & $4.21_{-2.58}^{+10.56}$ & $1.58_{-0.31}^{+0.63}$ & 1.19 & $18.73_{-0.20}^{+0.20}$ & $18.62_{-0.34}^{+0.24}$ \\\hline
    \end{tabularx}
    \egroup
    \caption{Halo properties of 8 classical dSphs based on  the Bayesian analysis adopting prior distribution using the Extended Press-Schechter model~\cite{Hiroshima:2018kfv}.}
    \label{tab:subhalos}
\end{table}

\section{Heavy dark matter model}
\label{s:dm_model}
We calculate the DM annihilation and decay spectra using {\tt HDMspectra}~\cite{Bauer:2020jay}. Ref.~\cite{Bauer:2020jay} achieves the matching between the scale of the DM mass (up to the Planck scale) and that below the electroweak scale. Ref.~\cite{Bauer:2020jay} also includes hadronization calculation and matches with the {\tt Pythia}~\cite{Sjostrand:2014zea}. In the scheme, the electroweak corrections are properly implemented. In this work, we consider DM masses $m_\chi$ ranging from $10^3$~GeV to $10^{11}$~GeV. We assume the DM particles annihilate or decay into a pair of Standard Model particles with a 100\% branching ratio. We consider 6 annihilation/decay channels: $b\Bar{b}$, $\mu^+\mu^-$, $\tau^+\tau^-$, $Z^0Z^0$, $W^+W^-$, and $h^0h^0$.

The energy fluxes of generated gamma rays, $E_\gamma^2 \Phi_\gamma^{(\rm gen)}(E_\gamma, \theta_\Delta)$, from DM annihilation/decay processes in the dSphs are:
\begin{equation}\label{eq:flux}
    E_\gamma^2 \Phi_\gamma^{(\rm gen)}(E_\gamma, \theta_\Delta)
    ={}{E}_\gamma^2\frac{dS_\gamma}{dE_\gamma}\times\left\{\begin{array}{ll}
         \frac{\langle\sigma v \rangle}{8\pi m_\chi^2} \times J^\mathrm{ann}(\theta_\Delta), & \mathrm{for\ annihilation,} \\
         \frac{1}{4\pi \tau_\chi m_\chi} \times J^\mathrm{dec}(\theta_\Delta), & \mathrm{for\ decay,}
    \end{array}\right.
\end{equation}
where $dS_\gamma/dE_\gamma$ is the gamma-ray spectra from DM annihilation/decay before accounting for electromagnetic cascades~\cite{Murase:2012xs}, $\langle\sigma v \rangle$ is the velocity-averaged cross section and $\tau_\chi$ is the life time of the decaying DM. The redshift dependence does not appear because of the proximity of the dSphs.

The spectra of generated gamma rays should be modified through electromagnetic cascades. We follow Ref.~\cite{Murase:2012rd} to calculate the cascade emission, approximating a galaxy to be a single zone. Details of electron-positron pair creation, synchrotron and inverse-Compton emission processes are considered. We solve the kinetic equation describing the evolution of the coupled system of photons and electrons~\cite{Murase:2012rd,Murase:2012xs}. For the magnetic field in target dSphs, we assume $B=1\ \mu$G as a fiducial value, while photons from the CMB~\cite{2009ApJ...707..916F} are taken as the background photon field for inverse-Compton emission. We also consider the infrared (IR) and optical radiation fields in dSphs, which are expected to be comparable to those in the clusters of galaxies~\cite{Lin:2003su, Kirby:2013wna}. Therefore, as an approximation, we include the low-IR extragalactic background light model in Ref.~\cite{Kneiske:2003tx} with 10 times enhancement {to account for contributions from the stellar populations in the dSphs} as in Refs.~\cite{Song:2023xdk, Murase:2012rd}. We ignore the spatial diffusion because high-energy electrons/positrons lose their energies faster than they diffuse out in a galaxy. {For the cascade calculation, we assume an escaping distance of the gamma rays at the $r_\mathrm{max}$ of the dSphs. This distance best represents the scale of the magnetic fields of the dSphs.} As a result, the expected gamma-ray energy fluxes at Earth, $E_\gamma^2 \Phi_\gamma(E_\gamma, \theta_\Delta)$, from DM annihilation/decay processes in the dSphs are expressed as:
\begin{equation}\label{eq:flux}
    E_\gamma^2 \Phi_\gamma(E_\gamma, \theta_\Delta)
    ={}
    {E}_\gamma^2\frac{dG_\gamma}{dE_\gamma}
    \times\left\{\begin{array}{ll}
         \frac{\langle\sigma v \rangle}{8\pi m_\chi^2} \times J^\mathrm{ann}(\theta_\Delta), & \mathrm{for\ annihilation,} \\
         \frac{1}{4\pi \tau_\chi m_\chi} \times J^\mathrm{dec}(\theta_\Delta), & \mathrm{for\ decay,}
    \end{array}\right.,
\end{equation}
where $dG_\gamma/dE_\gamma$ is the gamma-ray spectrum expected at Earth, which includes both attenuation and cascade components.
\begin{figure}
    \centering
    \includegraphics[width=0.48\columnwidth]{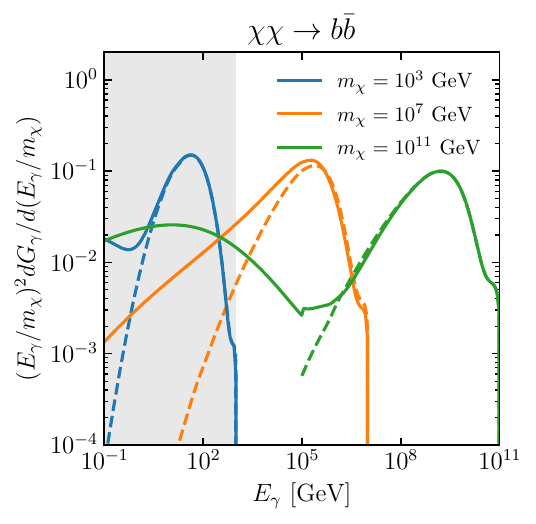}
    \includegraphics[width=0.48\columnwidth]{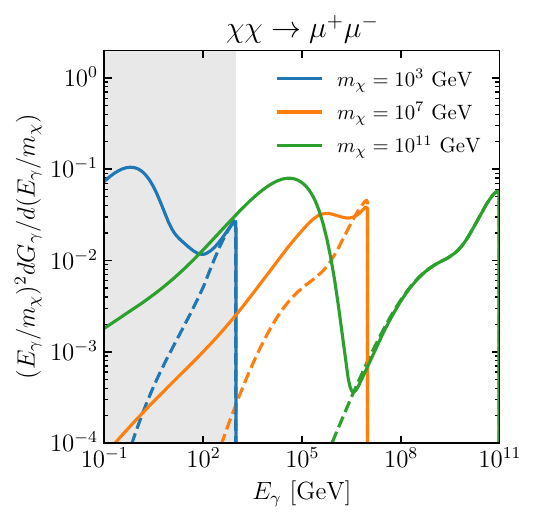}
    \caption{Gamma-ray spectra for annihilation of 3 DM masses: $m_\chi = 10^3$ GeV {(blue)}, $10^7$ GeV {(orange)}, and $10^{11}$ GeV {(green)}. {Solid lines show the expected spectra at Earth with electromagnetic cascades considered, while dashed lines show the generated spectra before accounting for the electromagnetic cascades.} The steep decline seen in the gamma-ray spectra corresponds to the kinematical cutoff by the DM mass.  {The gray bands indicate the energy range of the {\it Fermi}-LAT (100 MeV -- 1 TeV).}
    }
    \label{fig:spectra}
\end{figure}

Figure~\ref{fig:spectra} displays $dG_\gamma/dE_\gamma$ {(in solid lines)} originating from DM annihilation using our benchmark parameters. {The benchmark model assumes $B = 1\ \mu$G. It also includes the CMB and the extragalactic background light with 10 times enhancement.} We present the spectra for the $b\bar{b}$ channel in the left panel and the $\mu^+\mu^-$ channel in the right panel, considering three DM masses: $m_\chi = 10^3$ GeV, $10^7$ GeV, and $10^{11}$ GeV {(as shown respectively in blue, orange and green)}. The spectra are normalized for a single annihilation/decay event and are shown in $(E_\gamma/m_\chi)^2 dG_\gamma/d(E_\gamma/m_\chi)$ so that they are approximately the same level. The total fluxes from the target dSphs are determined by eq.~\ref{eq:flux} and are suppressed when $m_\chi$ increases. The gray bands in figure~\ref{fig:spectra} indicate the energy range of {\it Fermi}-LAT  (100 MeV -- 1 TeV). For heavy DM masses (e.g., $m_\chi = 10^7$ GeV and $m_\chi = 10^{11}$ GeV), the primary gamma-ray signals (which peak around $m_\chi$) are beyond the reach of {\it Fermi}-LAT. However, secondary emission extends to lower energies and still have sizable contributions to the {\it Fermi} energy range. {In figure~\ref{fig:spectra}, we also use dashed lines to show the generated spectra before accounting for the electromagnetic cascades.} 

\section{Data analysis}
\label{s:analysis}
We use the public software \texttt{fermipy}{~\cite{2017ICRC...35..824W}} to select the {\it Fermi}-LAT data, generate model templates convolved with the instrument response function, and perform the likelihood analysis. We select \texttt{P8R3\_SOURCE} events (with both \texttt{FRONT} and \texttt{BACK} types) in 14-year {\it Fermi}-LAT data (from Aug 4 2008 to Aug 4 2022) with energies from 100 MeV to 1 TeV. This event class provides an intermediate photon selection and is most suitable for moderately extended sources. We apply the standard quality filter \texttt{DATAQUAL>0\&\&LATCONFIG==1} and limit the maximum zentith angle to 90$^\circ$. For each dSph, the region of interest (ROI) is a $10^\circ\times 10^\circ$ squire centering the dSph. The data are binned into $0.1^\circ\times 0.1^\circ$ pixels and logarithmic energy bins with 5 bins per decade.

The expected photon count {$u^k_{ij}$} from the $i$-th pixel and $j$-th energy bin in the $k$-th dSph is
\begin{equation}
    u^k_{ij} = s^k_{ij}(A|m_\chi) + b^k_{ij}(\bm\lambda),
\end{equation}
where $s^k_{ij}$ and $b^k_{ij}$ are respectively the signal and the background counts from the $i$-th pixel and $j$-th energy bin in the $k$-th dSph. The signal $s^k_{ij}$ is determined by the DM model under consideration (including the subhalo model and electromagnetic cascades, see section~\ref{s:shmodel} and~\ref{s:dm_model}) and depends on the amplitude parameter $A$ (which is $\langle \sigma v \rangle$ for annihilation and $1/\tau_\chi$ for decay) for given $m_\chi$. We generate the DM spatial templates using the {\tt CLUMPY} package~\cite{Charbonnier:2012gf,Bonnivard:2015pia,Hutten:2018aix}, assuming the Navarro-Frenk-White profile~\cite{Navarro:1996gj} with truncation. See Appendix~\ref{sa:templates} for the details of the DM halo templates {before they are convolved with the {\it Fermi}-LAT Point Spread Function (PSF).} The background $b^k_{ij}$ includes all astrophysical emissions in the ROI and $\bm\lambda$ are the nuisance parameters for the background model. For the background components, we consider the Galactic diffuse emission\footnote{\texttt{gll\_iem\_v07.fits}}, the isotropic diffuse emission\footnote{\texttt{iso\_P8R3\_SOURCE\_V3\_v1.txt}}, and the resolved point sources in the 4FGL-DR4 catalog. The nuisance parameters include the normalization and spectral parameters of the Galactic and isotropic diffuse emissions and the point sources within $5^\circ$ from the dSphs. We also include point sources within a $20^\circ\times 20^\circ$ region centering the target dSph with their parameters fixed to the 4FGL values.

We adopt a joint Poisson likelihood function over all pixel $i$ and energy bin $j$ for the $k$-th dSph, which is
\begin{equation}
    \mathcal{L}^k(\bm\theta) = \prod_i \prod_j \dfrac{\mu^k_{ij}(\bm\theta)^{n^k_{ij}}e^{-\mu^k_{ij}(\bm\theta)}}{n^k_{ij}!}.
\end{equation}
Here, $n^k_{ij}$ and $\mu^k_{ij}(\bm\theta)$ are the observed and predicted photon counts at pixel $i$ and energy bin $j$ for the $k$-th dSph, respectively. In the meanwhile, $\bm\theta = \{A, \bm\lambda\}$ combine the DM and nuisance parameters. 

We use profile likelihood method~\cite{Rolke:2004mj} to derive constraints on $A$. The test statistics (TS) for any $A$ is defined as
\begin{equation}
    \mathrm{TS}(A) = -2\log\left(\dfrac{\mathcal{L}(A, \bm{\hat\lambda})}{\mathcal{L}(\bm{\hat\theta})}\right),
\end{equation}
where $\bm{\hat\theta}$ are the best-fitting parameters that maximize the likelihood function and $\bm{\hat\lambda}$ are the nuisance parameters that maximize the likelihood function for given $A$. The likelihood function $\mathcal{L}(\bm \theta)$ can be the likelihood function $\mathcal{L}^k(\bm \theta)$ of the $k$-th dSph when we put constraints from individual dSphs. Otherwise, we can use the total likelihood function of 8 dSphs, which is
\begin{equation}
    \mathcal{L}^\mathrm{total}(\bm \theta) = \prod_k \mathcal{L}^k(\bm\theta),
\end{equation}
to put constraints on $A$ from stacking 8 dSphs. In either case, the 95\% confidence level (CL) limit on $A$ is set by finding a $\mathrm{TS}= 2.71$.
 
\section{Results and discussion}
\label{s:results}

\begin{figure}[t!]
    \centering
    \includegraphics[width=0.48\columnwidth]{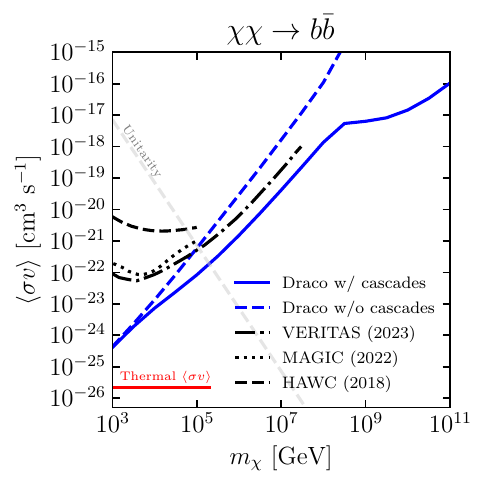}
    \includegraphics[width=0.48\columnwidth]{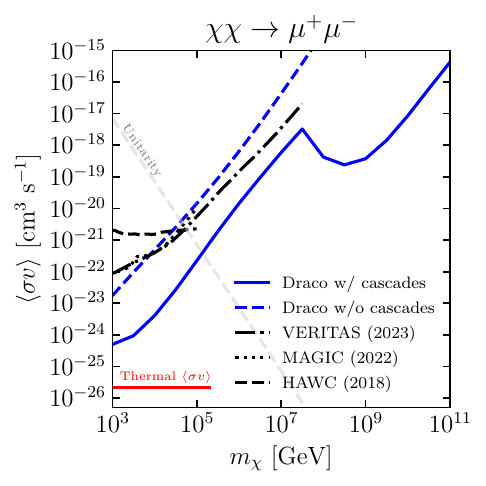}
    \caption{95\% CL upper limits on the annihilation cross section $\langle\sigma v\rangle$ for the $b\bar{b}$ (left panel) and $\mu^+\mu^-$ (right panel) channel from Draco. Constraints with (solid blue lines) and without (dashed blue lines) electromagnetic cascades are included. The constraints are compared with those from VERITAS~\cite{Acharyya:2023ptu} (dash dotted lines), MAGIC~\cite{MAGIC:2021mog} (dotted lines) and HAWC~\cite{HAWC:2017mfa} (dashed lines) with corrected J-factors (see section~\ref{s:results} for details). We also include the thermal cross section (up to $m_\chi \sim 200$ TeV, red line) and the partial wave unitarity bound for nearby dSphs (grey dashed line)~\cite{Tak:2022vkb, Smirnov:2019ngs}. 
    }
    \label{fig:limit_ann_Draco}
\end{figure}

In figure~\ref{fig:limit_ann_Draco}, we present the 95\% CL upper limits on heavy DM annihilation cross section ($\langle\sigma v\rangle$) from Draco for the $b\bar{b}$ (left panel) and $\mu^+\mu^-$ (right panel) channels. We consider two scenarios: constraints derived with (solid blue lines) and without (dashed blue lines) electromagnetic cascades. Without cascades, constraints for $m_\chi \gtrsim 10^8$~GeV with {\it Fermi}-LAT data are weaker than those in the literature~\cite{Acharyya:2023ptu, MAGIC:2021mog, HAWC:2017mfa}. The constraints are improved with {\it Fermi}-LAT by taking the cascade into consideration and in this work, we derive novel constraints for $m_\chi$ up to $10^{11}$ GeV. For the $b\bar{b}$ channel, constraints with and without electromagnetic cascades are nearly identical at $m_\chi \sim 10^3$ GeV, indicating that the constraints are primarily determined by the prompt gamma-ray emission around the WIMP mass scale. Constraints with cascades become stronger than those without cascades as $m_\chi$ increases because electromagnetic cascades dominate the signal in the {\it Fermi}-LAT energy range. In the case of the $\mu^+\mu^-$ channel, the annihilation is dominantly leptonic, and prompt gamma rays arise from final-state radiation. Therefore, the constraints get tighter by almost two orders of magnitude at $m_\chi \gtrsim 10^3$~GeV by analyzing with the cascade contribution. Overall, the constraints on the annihilation cross section $\langle\sigma v\rangle$ decrease as $m_\chi$ increases, as the annihilation rate is proportional to $1/m_\chi^2$. The feature at $m_\chi \sim 10^7 - 10^9$~GeV originates from the transition of the dominant processes of the secondary emission in the energy range of this analysis, from an inverse-Compton emission-dominated region to a synchrotron emission-dominated region. This change is more significant for the $\mu^+\mu^-$ channel (again due to its being dominantly leptonic), leading to a peak in the constraints around $10^9$ GeV. In figure~\ref{fig:limit_ann_Draco}, we also show limits with VERITAS~\cite{Acharyya:2023ptu} (dash dotted lines), MAGIC~\cite{MAGIC:2021mog} (dotted lines) and HAWC~\cite{HAWC:2017mfa} (dashed lines) constraints on the same dSph for comparison. To make the comparisons consistent, we rescale their constraints based on the J-factor used in this work, as listed in Table~\ref{tab:subhalos}. {We acknowledge that the linear scaling of the constraints ignores the discrepancies in the DM density profiles and PSFs of different instruments. Nonetheless, we anticipate these effects to be minor.} Our constraints with cascades are notably stronger than the constraints from those high-energy instruments in both channels. However, without cascades, our results are suppressed by VERITAS and MAGIC for $m_\chi \gtrsim 10^4$ GeV. In figure~\ref{fig:limit_ann_Draco}, we also show the thermal cross section up to $m_\chi \sim 200$ TeV~\cite{Smirnov:2019ngs, Tak:2022vkb} and the partial wave unitarity bound {for point-like particle DM} assuming the relative velocity $v_{\rm rel} \sim 2 \times 10^{-5}$ in the dSphs~\cite{Tak:2022vkb}.

\begin{figure}[h!]
    \centering
    \includegraphics[width=0.48\columnwidth]{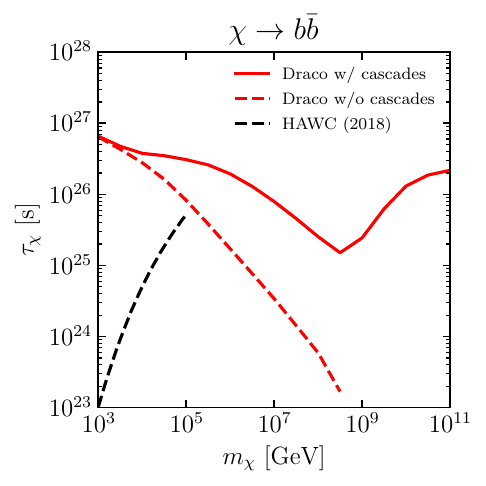}
    \includegraphics[width=0.48\columnwidth]{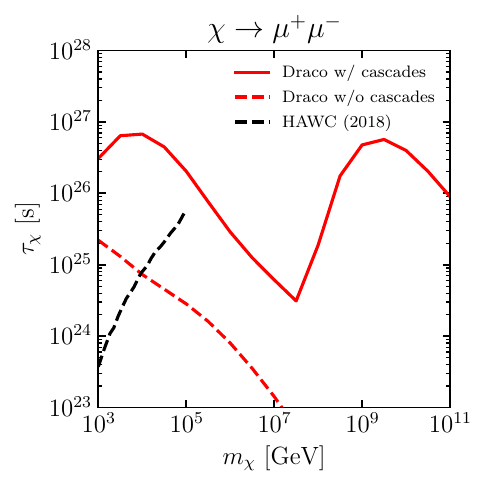}
    \caption{95\% CL lower limits on the lifetime $\tau_\chi$ for the $b\bar{b}$ (left pannel) and $\mu^+\mu^-$ (right pannel) channel from Draco. Constraints with (solid red lines) and without (dashed red lines) electromagnetic cascades are included.}
    \label{fig:limit_dec_Draco}
\end{figure}

Figure~\ref{fig:limit_dec_Draco} displays the 95\% CL lower limits on heavy DM lifetime ($\tau_\chi$). Similar to the annihilation case, the inclusion of the electromagnetic cascades significantly improves the constraints for heavy $m_\chi$, particularly for the $\mu^+\mu^-$ channel starting from $m_\chi\sim 10^3$ GeV. Unlike the annihilation case, constraints on the lifetime $\tau$ are only mildly weakened with increasing $m_\chi$ from $10^3$ GeV, as the decay rate is proportional to $1/m_\chi$. Synchrotron emission also tightens the constraints on $\tau$ for $m_\chi \gtrsim 10^7 - 10^9$ GeV.

\begin{figure}[t!]
    \centering
    \includegraphics[width=0.32\columnwidth]{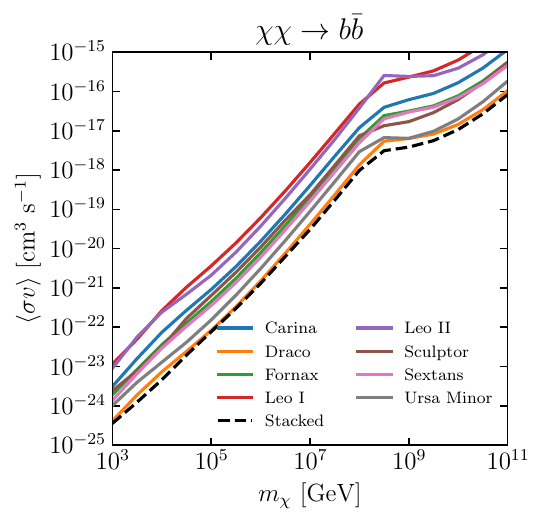}
    \includegraphics[width=0.32\columnwidth]{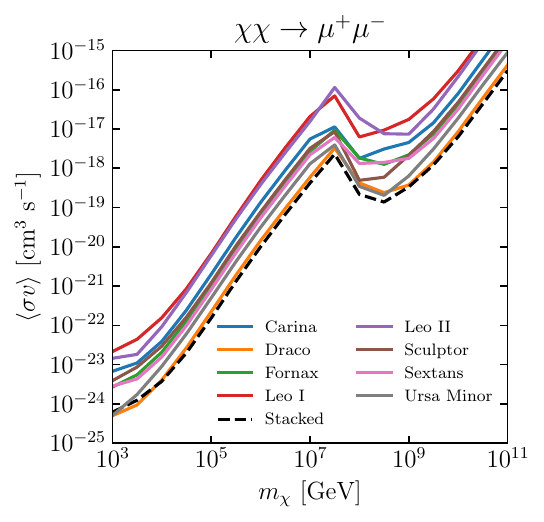}
    \includegraphics[width=0.32\columnwidth]{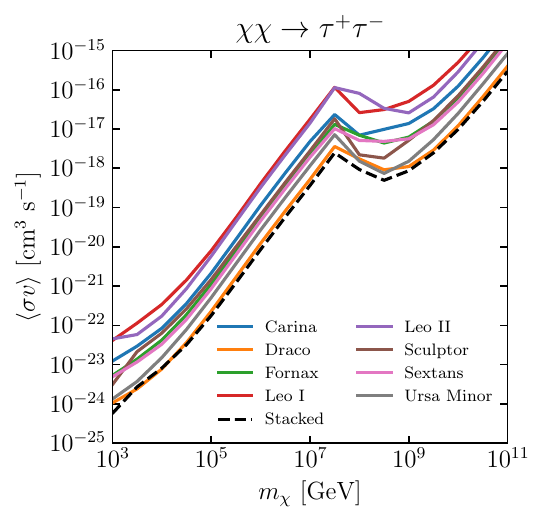}
    \includegraphics[width=0.32\columnwidth]{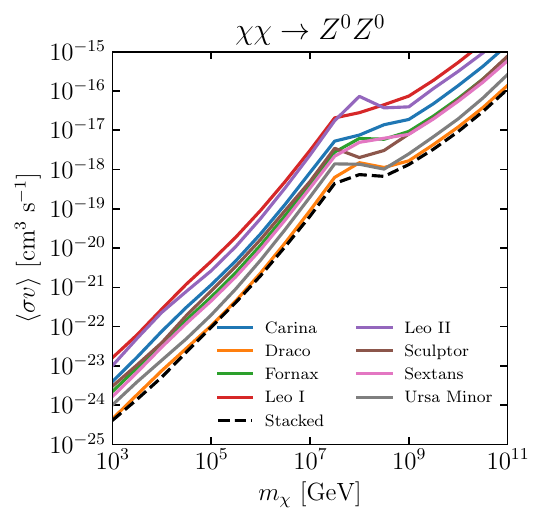}
    \includegraphics[width=0.32\columnwidth]{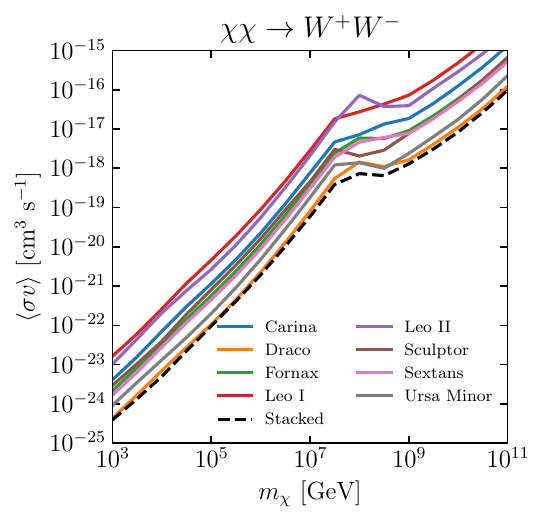}
    \includegraphics[width=0.32\columnwidth]{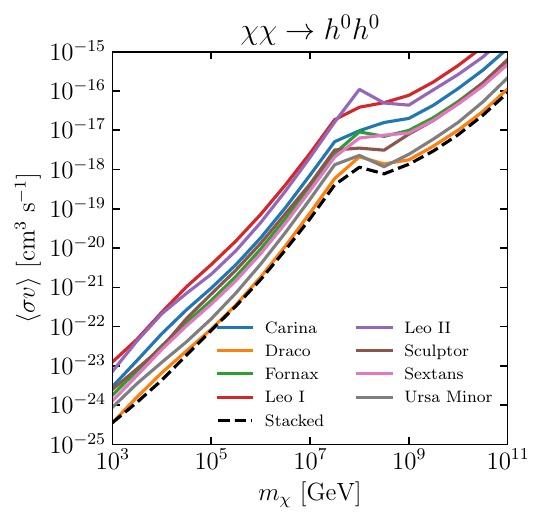}
    \caption{95\% CL upper limits on the annihilation cross section $\langle\sigma v\rangle$ for 6 channels and 8 dSphs. We also include the limits from stacking 8 dSphs (dashed lines).}
    \label{fig:limits_ann}
\end{figure}

\begin{figure}[h!]
    \centering
    \includegraphics[width=0.32\columnwidth]{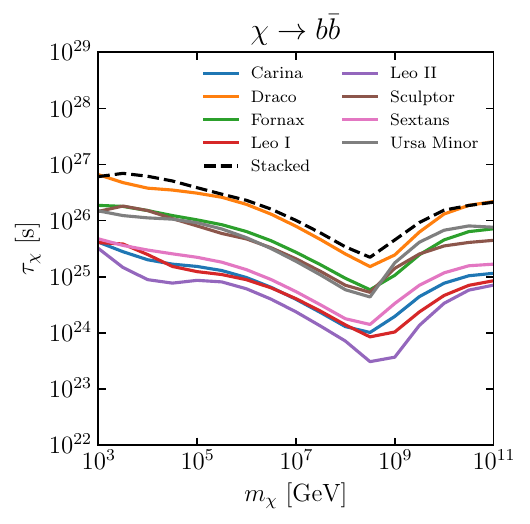}
    \includegraphics[width=0.32\columnwidth]{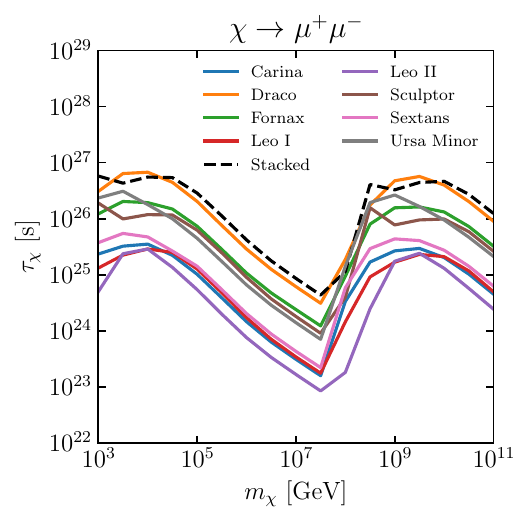}
    \includegraphics[width=0.32\columnwidth]{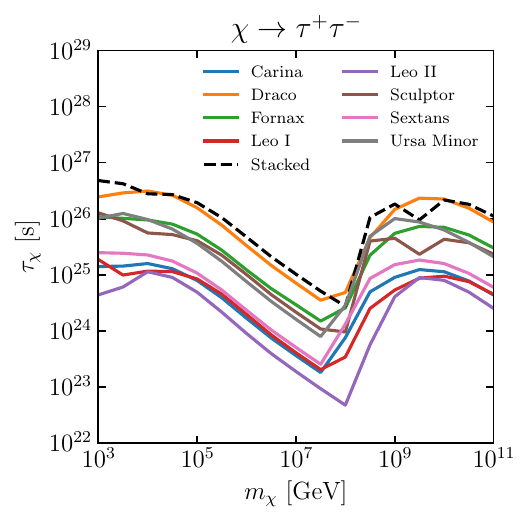}
    \includegraphics[width=0.32\columnwidth]{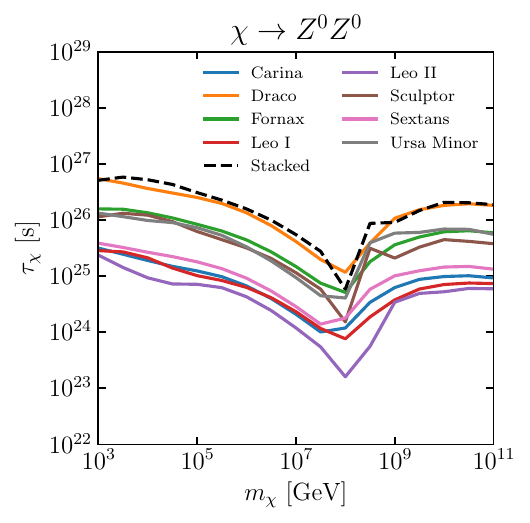}
    \includegraphics[width=0.32\columnwidth]{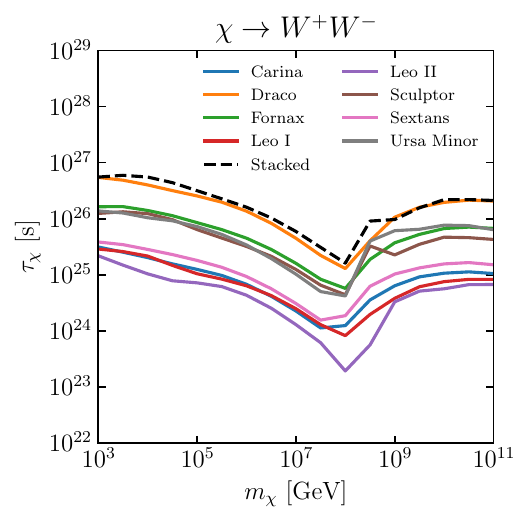}
    \includegraphics[width=0.32\columnwidth]{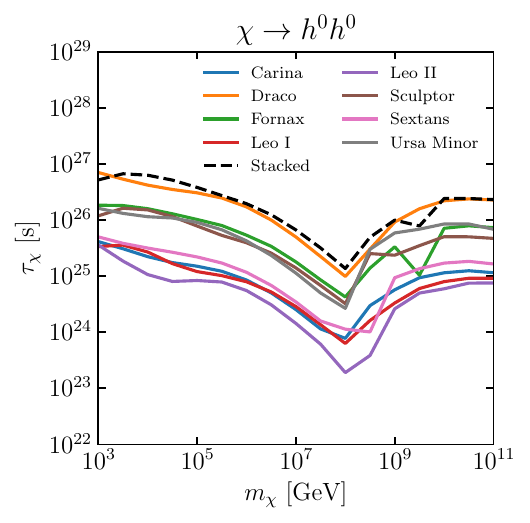}
    \caption{95\% CL lower limits on the lifetime $\tau_\chi$ for 6 channels and 8 dSphs. We also include the limits from stacking 8 dSphs (dashed lines).}
    \label{fig:limits_dec}
\end{figure}

We present a complete set of constraints on heavy DM annihilation cross section and decay lifetime in figure~\ref{fig:limits_ann} and figure~\ref{fig:limits_dec}, respectively. These constraints encompass six annihilation/decay channels and are derived from 8 classical dSphs. Among these dSphs, Draco stands out as the most stringent individual constraint for both annihilation and decay cases due to its substantial J-factors. The structures in constraints for $m_\chi \gtrsim 10^7 - 10^9$ GeV, attributed to synchrotron emission, are a consistent feature across all dSphs and channels. This effect is more pronounced for channels involving leptons, such as $\mu^+\mu^-$ and $\tau^+\tau^-$. In addition to individual dSph constraints, we also include constraints derived from stacking all 8 dSphs, as described in the section~\ref{s:analysis}. We observe slightly improved limits from stacking 8 dSphs compared to the strongest limits from individual dSphs. {The stacking limits are most likely driven by Draco, which has the largest J-factors among the target dSphs.}

\begin{figure}[t!]
    \centering
    \includegraphics[width=0.48\columnwidth]{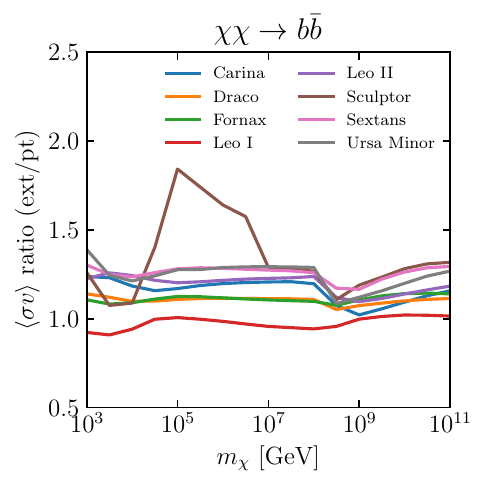}
    \includegraphics[width=0.48\columnwidth]{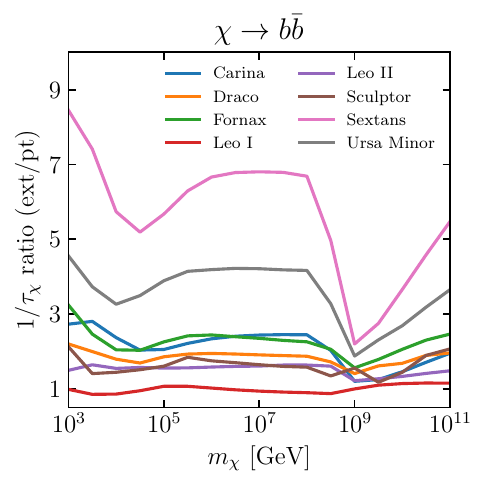}
    \caption{Ratios of the constraints on $\langle\sigma v\rangle$ (left panel) and $1/\tau_\chi$ (right panel)  between the extended and point-source analysis. The annihilation/decay channel is $b\bar{b}$ and 8 dSphs are included.}
    \label{fig:limit_ext_vs_ps}
\end{figure}

As a baseline of the analysis, we have assumed that dSphs are extended sources, and their DM halos follow the NFW density profile with truncation. A recent study~\cite{DiMauro:2022hue} investigated the impact of considering the extension of dSphs when searching for DM signals with {\it Fermi}-LAT. They found that modeling dSphs as extended sources weakened the annihilation constraints by a factor of approximately 2, depending on the specific dSph and channel under consideration. To explore this effect, we treat the dSphs as point sources and repeat the data analysis for the $b\bar{b}$ channel. We make the assumption that the J-factors of the dSphs in the point-source approximation are equal to the $J^\mathrm{ann}(\theta_\Delta)$ and $J^\mathrm{dec}(\theta_\Delta)$ as presented in Table~\ref{tab:subhalos} for the extended cases, which corresponds to {an aggressive assumption for the point-source hypothesis}. Figure~\ref{fig:limit_ext_vs_ps} shows the ratios of the constraints on $\langle\sigma v\rangle$ (left panel) and $1/\tau_\chi$ (right panel) between the extended and point-source analyses for the $b\bar{b}$ channel. We observe that the constraints are generally weakened in the extended analysis. The ratios for the annihilation $\langle\sigma v\rangle$ vary from approximately 1 to 2, depending on the target dSph and mass $m_\chi$. The range of ratios for $1/\tau_\chi$ is more variable, ranging from around 1 to nearly 9. The weakening effect is more pronounced for DM decay since the extended signal for DM decay is less concentrated at the center of the dSphs, as demonstrated in the 2D templates in Appendix~\ref{sa:templates}. Our results are qualitatively consistent with the findings of Ref.~\cite{DiMauro:2022hue}.

\begin{figure}[h!]
    \centering
    \includegraphics[width=0.48\columnwidth]{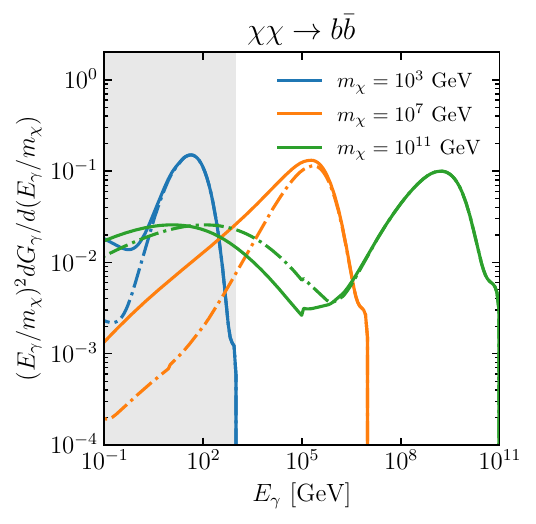}
    \includegraphics[width=0.48\columnwidth]{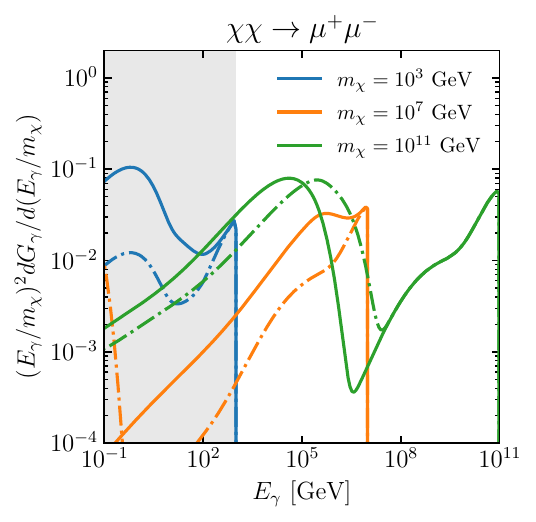}
    \caption{{Gamma-ray spectra from DM annihilation with 3 different masses: $m_\chi = 10^3$ GeV {(blue)}, $10^7$ GeV {(orange)}, and $10^{11}$ GeV {(green)}. The solid lines consider $B = 1\ \mu$G, the fiducial value for the main results. The dash-dotted lines consider an alternative magnetic field strength with $B = 10\ \mu$G.}}
    \label{fig:spectra_B}
\end{figure}

\begin{figure}[h!]
    \centering
    \includegraphics[width=0.48\columnwidth]{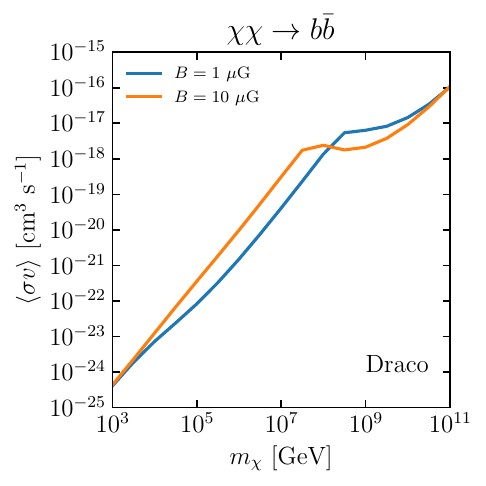}
    \includegraphics[width=0.48\columnwidth]{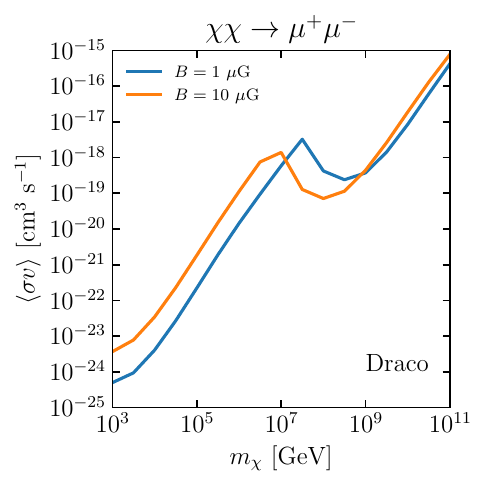}
    \caption{{Constraints on $\langle\sigma v\rangle$ for the $b\bar{b}$ (left panel) and $\mu^+\mu^-$ (right panel) channels using Draco as dSph target. The plots show the comparison between $B=1\ \mu$G and $B=10\ \mu$G respectively in blue and orange.}
    }
    \label{fig:limit_ann_B}
\end{figure}

\begin{figure}[h!]
    \centering
    \includegraphics[width=0.48\columnwidth]{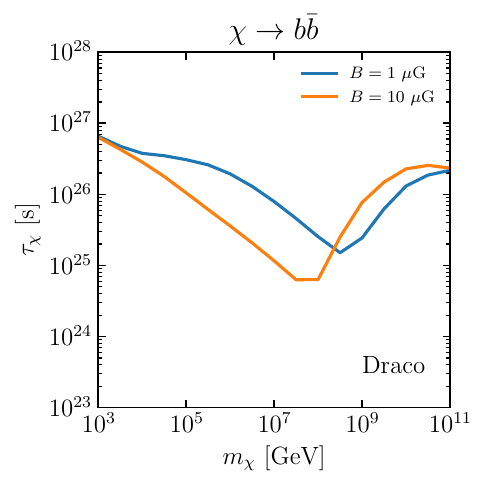}
    \includegraphics[width=0.48\columnwidth]{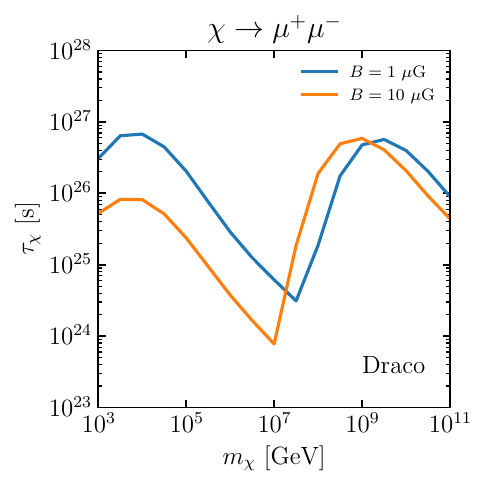}
    \caption{{Constraints on $\tau_\chi$ for the $b\bar{b}$ (left panel) and $\mu^+\mu^-$ (right panel) channels using Draco as dSph target. The plots show the comparison between $B=1\ \mu$G and $B=10\ \mu$G respectively in blue and orange.
    }
    }
    \label{fig:limit_dec_B}
\end{figure}

Finally, we investigate the impact of varying magnetic fields in the dSphs. The magnetic field strength could alter the results since the dominant secondary process in the {\it Fermi} energy range is affected by the energy partition between the background magnetic field and photon field. Ref.~\cite{2011A&A...529A..94C} has shown that the magnetic field in dSphs are usually less than a few $\mu$G and can reach values as high as $\sim 10\ \mu$G. In our fiducial model, we assume a magnetic field strength of $B = 1\ \mu$G. 
{The energy density of the magnetic field strength in this range is similar to that of the CMB. More specifically, a magnetic field with $B = 3.24\ \mu$G has the same energy density of the CMB~\cite{Dolag:2004kp}. Varying magnetic field in this range is important as the energy losses of the inverse-Compton and synchrotron processes are comparable.}
Therefore, we test $B = 10\ \mu$G to reflect an extreme case and consider systematics coming from the magnetic field variation within a galaxy. 
{Figure~\ref{fig:spectra_B} displays the expected gamma-ray spectra at Earth for the $b\bar{b}$ channel in the left panel and the $\mu^+\mu^-$ channel in the right panel for three DM masses, similar to figure~\ref{fig:spectra}. The solid lines correspond to $B = 1\ \mu$G, the fiducial value for the main results, while the dash-dotted lines represent an alternative magnetic field strength of $B = 10\ \mu$G.}
Figure~\ref{fig:limit_ann_B} compares the constraints on $\langle\sigma v\rangle$ between $B = 1\ \mu$G and $B = 10\ \mu$G for the $b\bar{b}$ (left panel) and $\mu^+\mu^-$ (right panel) channels. Generally, an increase in the magnetic field reduces the inverse-Compton emission and increases the synchrotron emission. Therefore, constraints with $B = 10\ \mu$G are weaker than those with $B = 1\ \mu$G for $m_\chi\lesssim10^7-10^8$~GeV in which the inverse-Compton emission is more important than the synchrotron emission in the LAT energy range, and vice versa. In most cases, the constraints change by less than one order of magnitude, depending on $m_\chi$. In figure~\ref{fig:limit_dec_B}, we make the same comparison for constraints on $\tau_\chi$. Once again, changing $B$ from $1\ \mu$G to $10\ \mu$G alters the constraints by at most one order of magnitude.

\section{Summary}
\label{s:conclusion}
We used 14 years of {\it Fermi}-LAT data to search for gamma-ray signals from heavy DM with $m_\chi\gtrsim 10^3$~GeV in 8 classical dSphs, and constrained the annihilation cross section and decay lifetime.
In particular, we incorporated the effects of electromagnetic cascades to better probe DM heavier than 1~TeV together with the spatial extension of target dSphs considering their cosmological evolution under the gravitational potential of the Milky Way. 
We also quantified the impacts of the spatial extension and the magnetic field strength, and found that resulting systematic uncertainties are less than one order of magnitude. 
We showed that our dSph constraints from the LAT non-detection of gamma-ray signals with electromagnetic cascades surpass not only those without cascades but also the limits derived from very high-energy gamma-ray facilities such as VERITAS, MAGIC and HAWC. Our findings offer valuable complementary constraints on heavy DM, in conjunction with observations of high-energy gamma rays (e.g., from galaxy clusters~\cite{Murase:2012rd,Song:2023xdk} and the Milky Way halo~\cite{Cohen:2016uyg}), cosmic rays~\cite{Ishiwata:2019aet,Das:2023wtk}, and neutrinos~\cite{Murase:2012xs,Chianese:2021htv}.

We demonstrated that incorporating the electromagnetic cascades  reinforces the dSph search for heavy DM by {\it Fermi}-LAT. Future observations of gamma rays and neutrinos will also benefit from accounting for such effects~\cite{Leung:2023gwp}. Meanwhile, it is crucial to consider the cosmological evolution of dSphs in the Milky Way to accurately estimate their J-factors and spatial extensions, which will enable us to establish more reliable constraints~\cite{Crnogorcevic:2023ijs} or possibly even detection~\cite{McDaniel:2023bju} in the future.

\acknowledgments

{The \textit{Fermi} LAT Collaboration acknowledges generous ongoing support
from a number of agencies and institutes that have supported both the
development and the operation of the LAT as well as scientific data analysis.
These include the National Aeronautics and Space Administration and the
Department of Energy in the United States, the Commissariat \`a l'Energie Atomique
and the Centre National de la Recherche Scientifique / Institut National de Physique
Nucl\'eaire et de Physique des Particules in France, the Agenzia Spaziale Italiana
and the Istituto Nazionale di Fisica Nucleare in Italy, the Ministry of Education,
Culture, Sports, Science and Technology (MEXT), High Energy Accelerator Research
Organization (KEK) and Japan Aerospace Exploration Agency (JAXA) in Japan, and
the K.~A.~Wallenberg Foundation, the Swedish Research Council and the
Swedish National Space Board in Sweden.
 
Additional support for science analysis during the operations phase is gratefully 
acknowledged from the Istituto Nazionale di Astrofisica in Italy and the Centre 
National d'\'Etudes Spatiales in France. This work performed in part under DOE 
Contract DE-AC02-76SF00515.}

{The authors express our gratitude to Soheila Abdollahi, Regina Caputo, Milena Crnogorcevic, and Davide Serini for their valuable assistance in preparing the draft. We especially thank  Davide Serini, Regina Caputo, and Donggeun Tak for providing insightful comments on the draft.}
We also thank Shin'iciro Ando for sharing the codes on the DM template model for the dwarf satellites. 
D.S., N.H., and K.M. are supported by JSPS KAKENHI Grant Number 20H05852. 
The work of N.H is also supported in part by the Scientific Research 22K14035 MEXT Leading Initiative for
Excellent Young Researchers Grant Number 2023L0013.  
The work of K.M. was also supported by the NSF Grants No.~AST-2108466 and No.~AST-2108467, and KAKENHI No.~20H01901.

\paragraph{Note added.} 
While we are finalizing this paper, Ref.~\cite{Hu:2023iex} of a close research interest appeared. The major difference is the treatment of the source extension. Also, the model for the diffusion is different from each other.

\appendix
\section{Spatial templates}
\label{sa:templates}
In our data analysis, we model dSphs as extended sources. The spatial template is determined by median values of $\rho_s$ and $r_s$ as described in section~\ref{s:shmodel}. We use the {\sc CLUMPY} package to calculate the differential J-factors $\frac{dJ^\mathrm{ann}}{d\Omega}(\Omega)$ and $\frac{dJ^\mathrm{dec}}{d\Omega}(\Omega)$ over the ROIs of dSphs. 

\begin{figure}[t!]
    \centering
\includegraphics[width=0.24\textwidth]{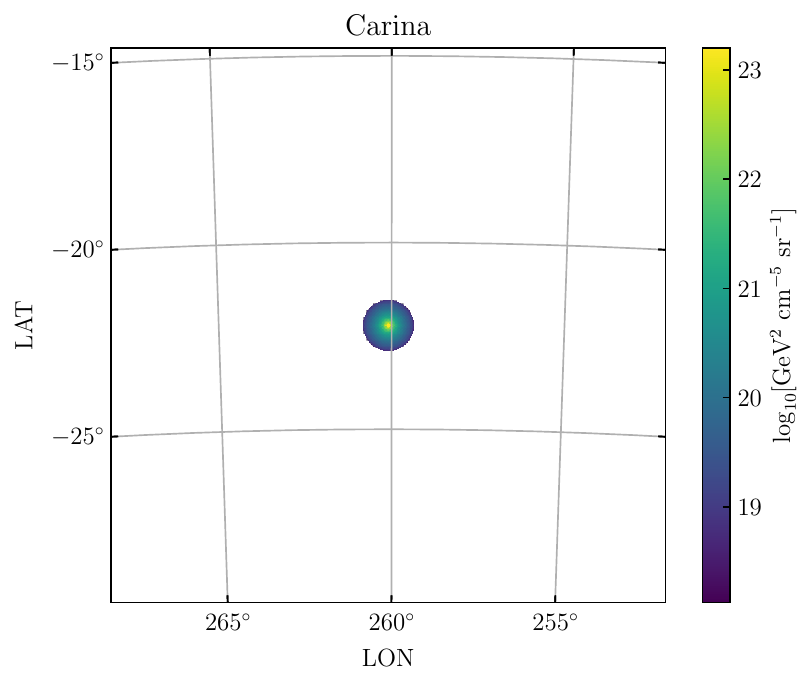}
\includegraphics[width=0.24\textwidth]{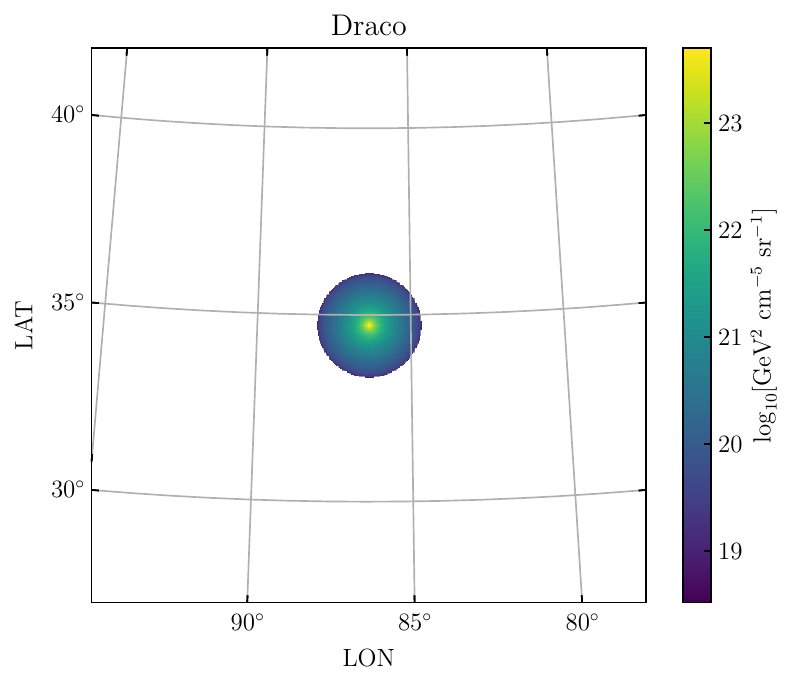}
\includegraphics[width=0.24\textwidth]{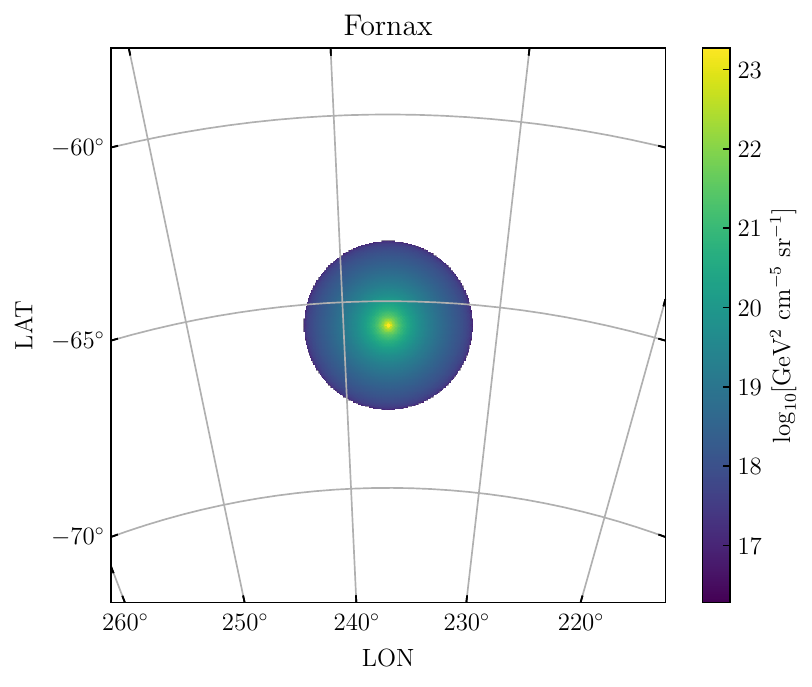}
\includegraphics[width=0.24\textwidth]{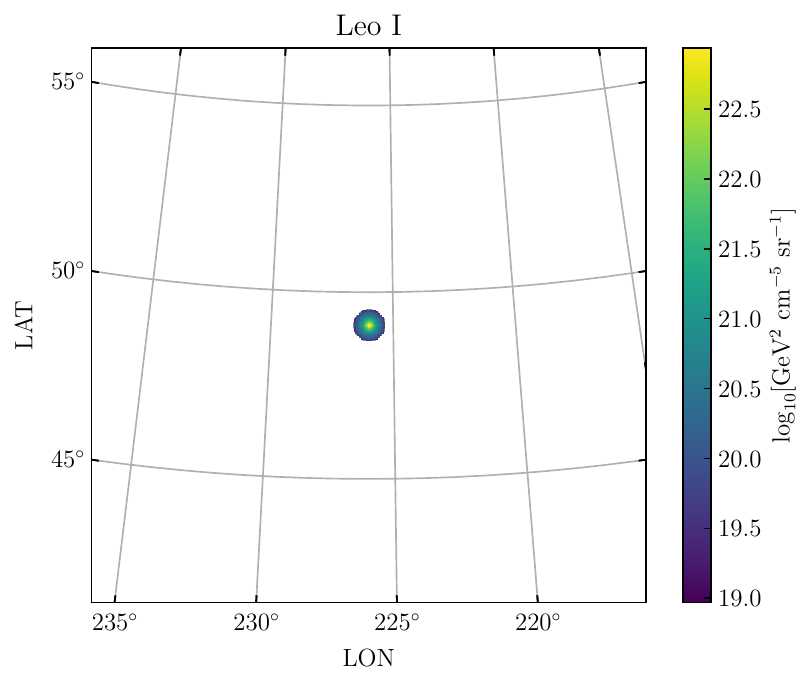}
\includegraphics[width=0.24\textwidth]{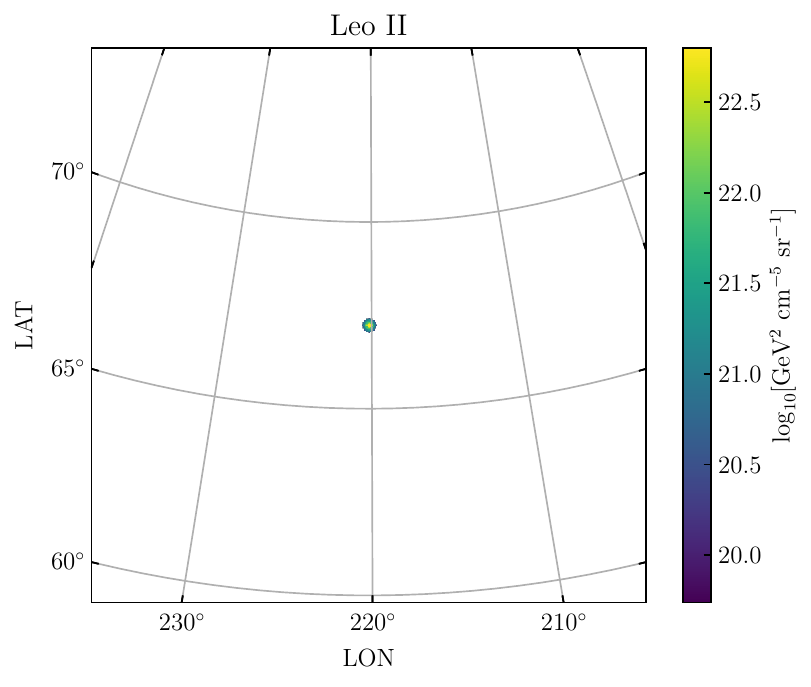}
\includegraphics[width=0.24\textwidth]{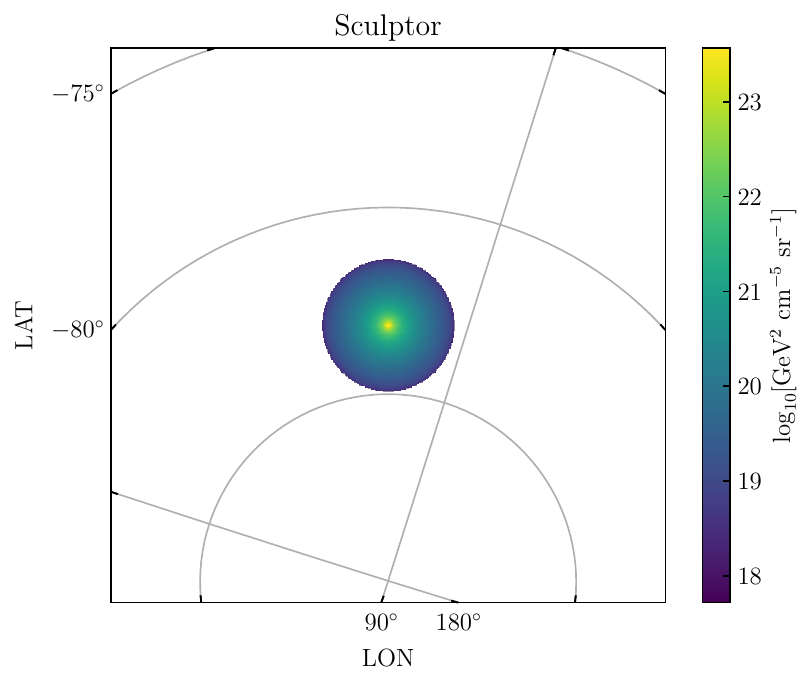}
\includegraphics[width=0.24\textwidth]{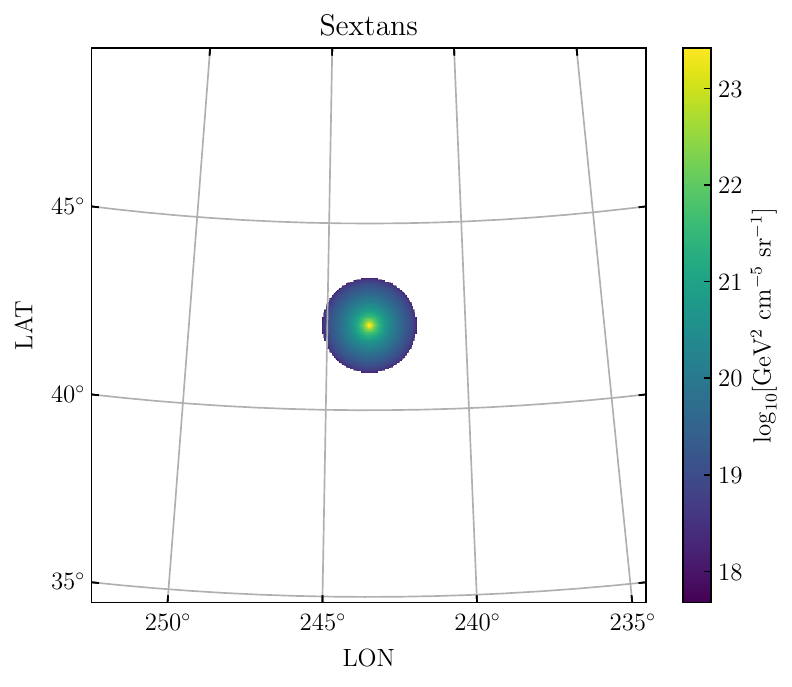}
\includegraphics[width=0.24\textwidth]{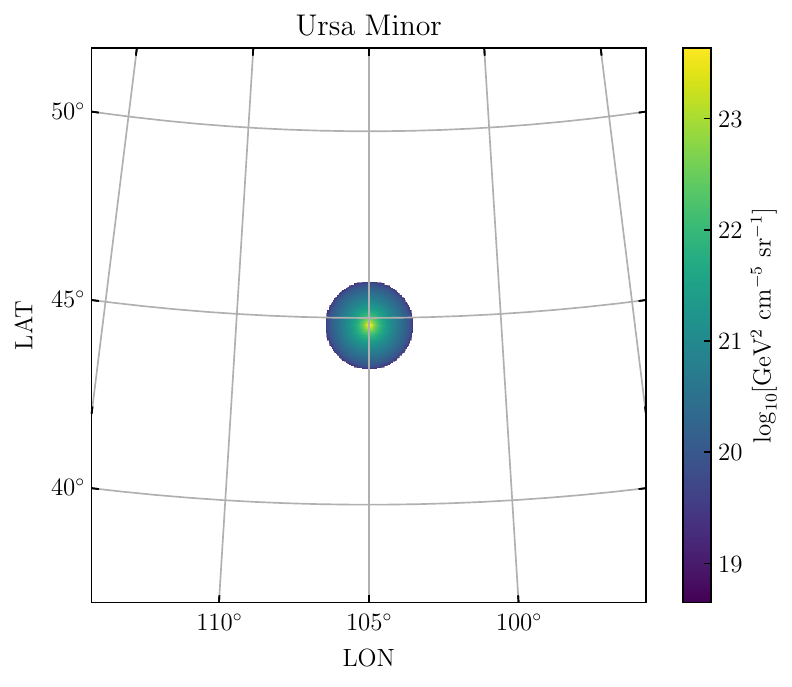}
    \caption{$\frac{dJ^\mathrm{ann}}{d\Omega}(\Omega)$ for 8 dSphs.
     }
    \label{fig:Jfactor}
\end{figure}

\begin{figure}[h!]
    \centering
\includegraphics[width=0.24\textwidth]{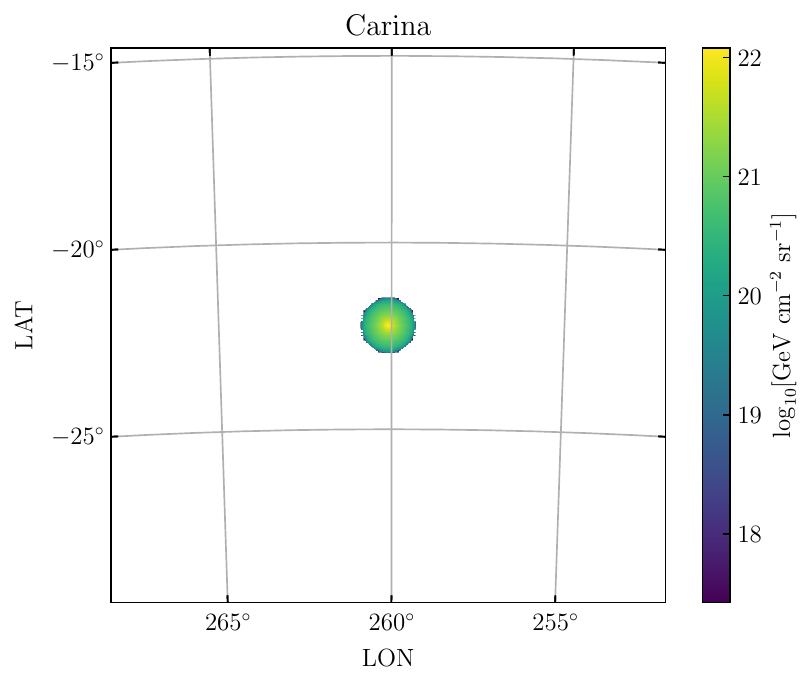}
\includegraphics[width=0.24\textwidth]{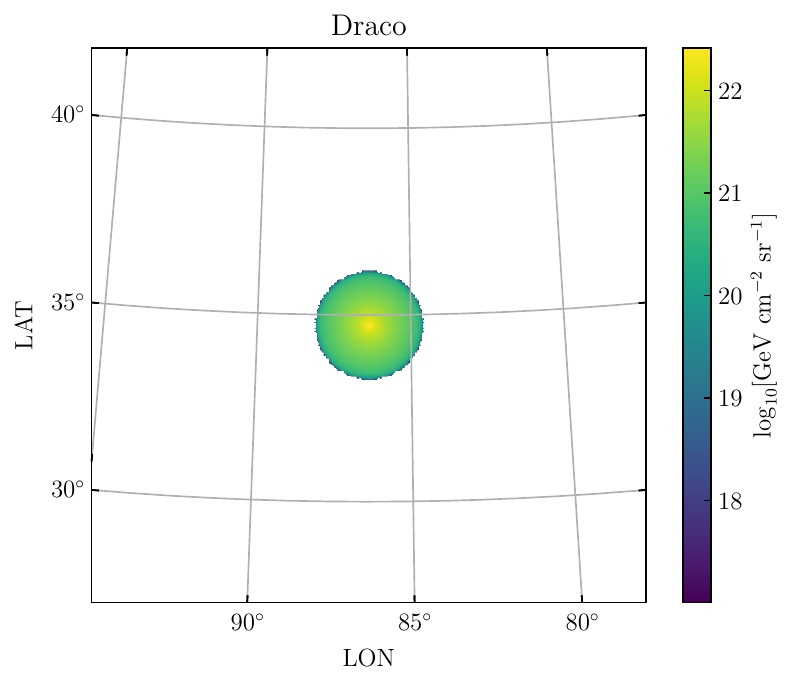}
\includegraphics[width=0.24\textwidth]{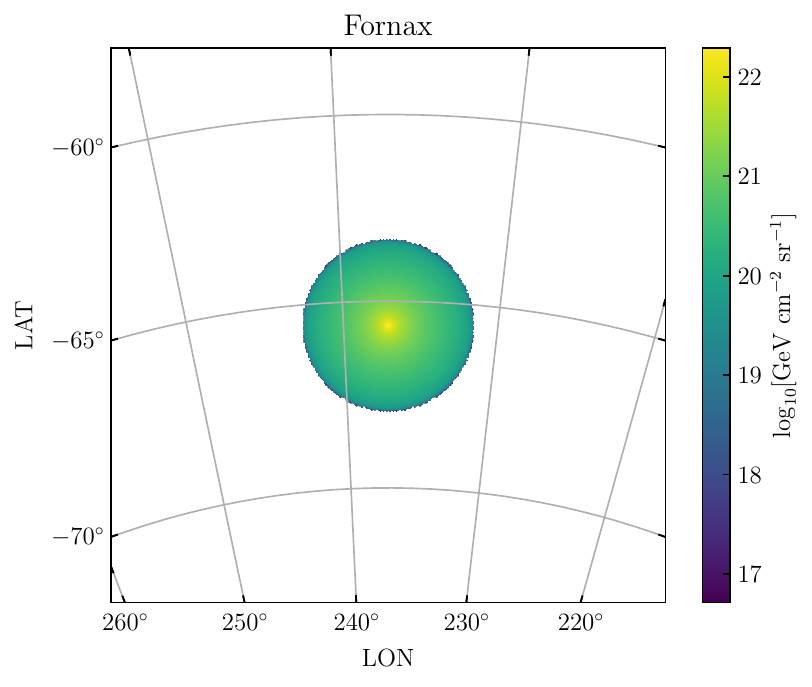}
\includegraphics[width=0.24\textwidth]{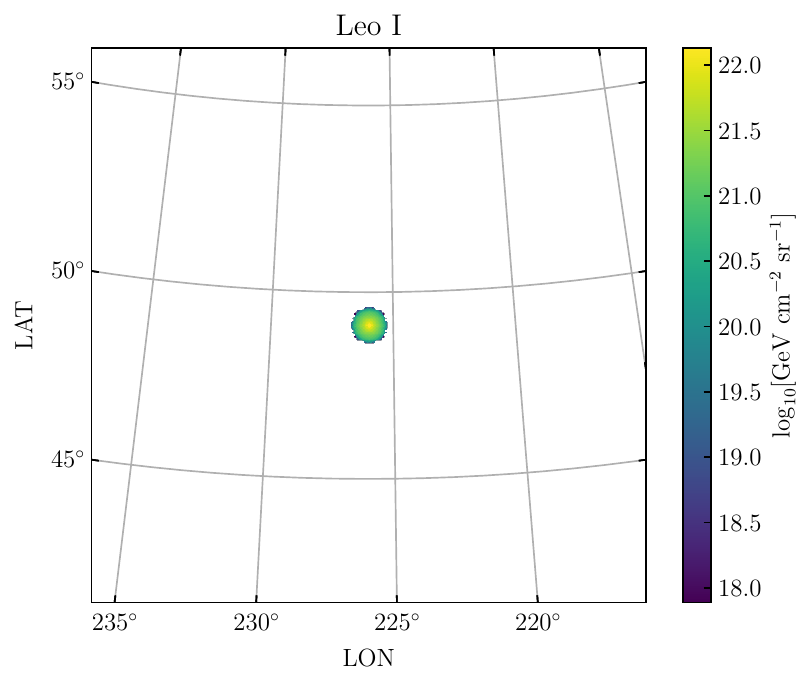}
\includegraphics[width=0.24\textwidth]{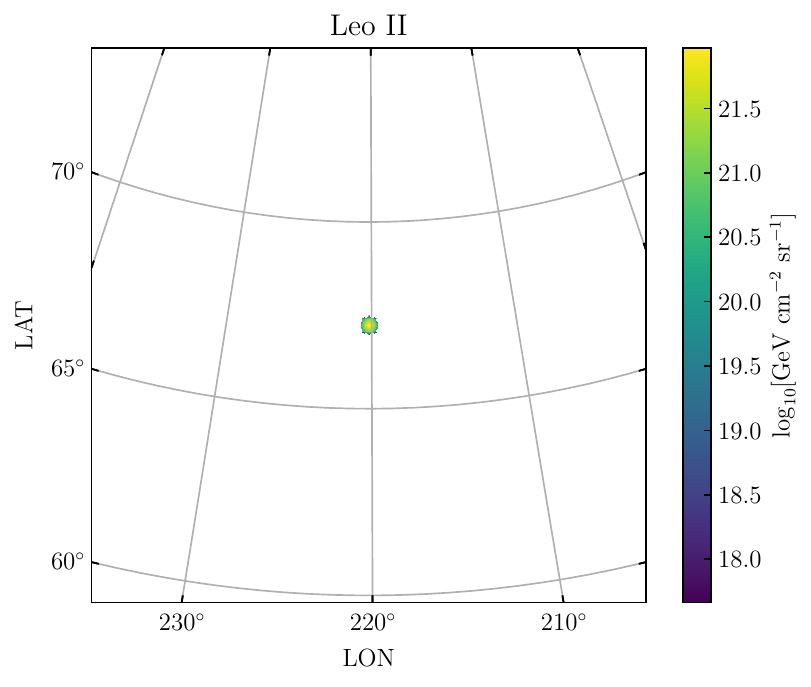}
\includegraphics[width=0.24\textwidth]{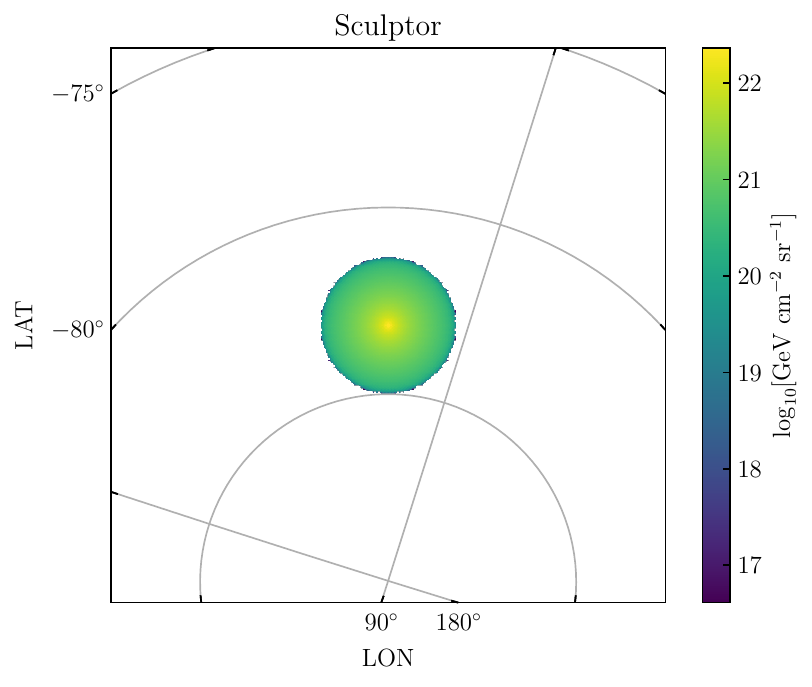}
\includegraphics[width=0.24\textwidth]{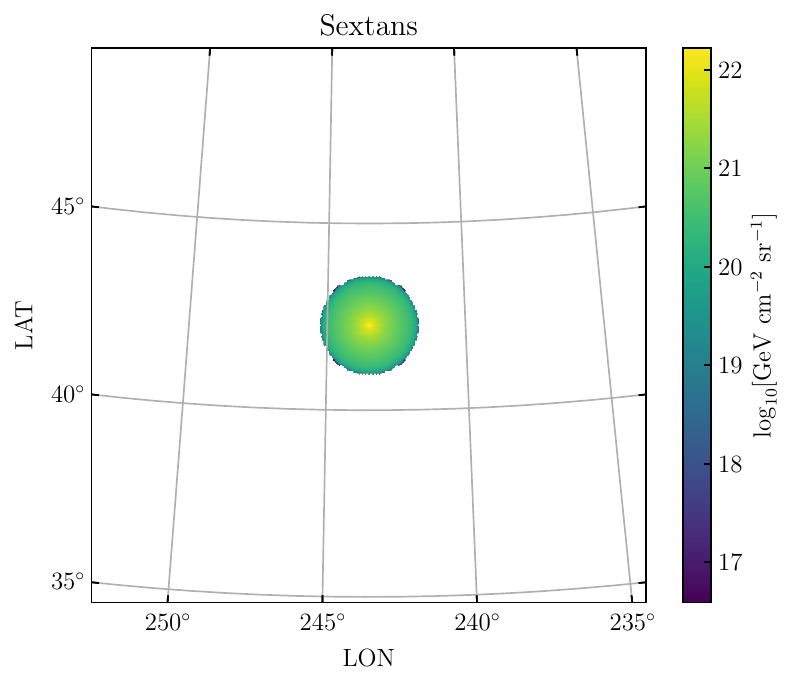}
\includegraphics[width=0.24\textwidth]{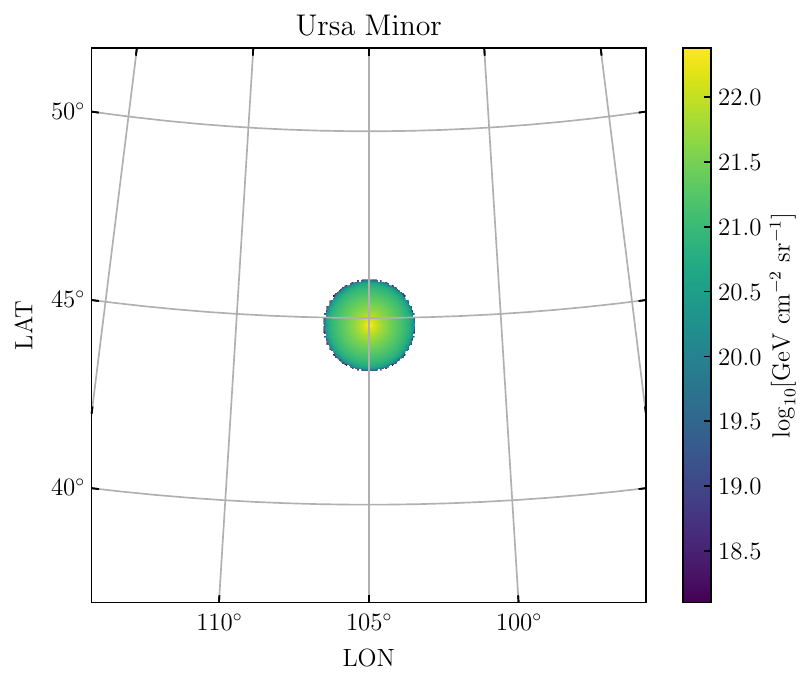}
    \caption{$\frac{dJ^\mathrm{dec}}{d\Omega}(\Omega)$ for 8 dSphs.}
    \label{fig:Dfactor}
\end{figure}

Figure~\ref{fig:Jfactor} shows the templates for $\frac{dJ^\mathrm{ann}}{d\Omega}(\Omega)$ for 8 target dSphs, {before they are convolved with the {\it Fermi}-LAT PSF.} We calculate $\frac{dJ^\mathrm{ann}}{d\Omega}(\Omega)$ up to the $\theta_\Delta$ for each dSph (see Table~\ref{tab:subhalos}), while figure~\ref{fig:Dfactor} shows the templates for $\frac{dJ^\mathrm{dec}}{d\Omega}(\Omega)$. In the case of annihilation, the DM signals are highly concentrated at the centers of the dSphs. However, in the case of decay, the DM signals are less concentrated and more diffuse, evenly distributed.


\bibliographystyle{jhep}
\bibliography{main}

\end{document}